\documentstyle[12pt,color,epsfig,subfigure]{article}
\advance\textheight by \topskip
\newcommand{\comment}[1]{}

\def\met	{\not \! E_T }

\def\wt		{\widetilde}
\def\hstop	{\wt{t}_{2}}
\def\lstop	{\wt{t}_{1}}
\def\stop	{\wt{t}}
\def\mlstop	{m_{\lstop}}
\def\mstop	{m_{\stop}}
\def\lspone	{\wt\chi_1^0}
\def\mlspone	{m_{\lspone}}

\def\gev	{\,GeV}
\def\tt		{t\bar{t}}
\def\bb		{b\bar{b}}
\def\cc		{c\bar{c}}
\def\stopl      {\wt{t}_{L}}
\def\stopr      {\wt{t}_{R}}

\def\delm       {\mlstop - \mlspone}
\def\Del        {\Delta}
\def\ra         {\rightarrow}

\def\loopdk{\lstop \ra c \lspone}

\def\beq	{\begin{equation}}
\def\eeq	{\end{equation}}
\def\beqn	{\begin{eqnarray}}
\def\eeqn	{\end{eqnarray}}
\def\issue(#1,#2,#3){{\bf #1} (#2) #3}
\def\PRD(#1,#2,#3){Phys. Rev. D \issue(#1,#2,#3)}
\def\PLB(#1,#2,#3){Phys. Lett. B  \issue(#1,#2,#3)}
\def\PRL(#1,#2,#3){Phys. Rev. Lett. \issue(#1,#2,#3)}
\def\PREP(#1,#2,#3){Phys.\ Rep. \issue(#1,#2,#3)}
\def\NPB(#1,#2,#3){Nucl.\ Phys.\ B \issue(#1,#2,#3)}
\def\JHEP(#1,#2,#3){J.\ High\ Energy\ Phys. \issue(#1,#2,#3)}
\def\PTP(#1,#2,#3){Prog.\ Theor.\ Phys. \issue(#1,#2,#3)}
\def\CPC(#1,#2,#3){Comp.\ Phys.\ Comm. \issue(#1,#2,#3)}
\def\ASTR(#1,#2,#3){Astropart.\ Phys. \issue(#1,#2,#3)}
\begin{document}
\tolerance=100000
\thispagestyle{empty}
\setcounter{page}{0}
\vspace*{\fill}
\vspace{-0.5in}
\begin{flushright}
IISER/HEP/08/07
\end{flushright}
\begin{center}
{\Large \bf
Search for Top Squarks at Tevatron Inspired by Dark Matter and Electroweak 
Baryogenesis}\\[1.00cm]
{\large Nabanita Bhattacharyya$^{a,}$,\footnote{\it nabanita@iiserkol.ac.in}
        Amitava Datta$^{a,}$,\footnote{\it adatta@iiserkol.ac.in}
        and Manas Maity$^{b,}~$\footnote{\it manas.maity@cern.ch}\\
$^a$ {\it  Indian Institute of Science Education and Research, Kolkata, Salt Lake City,
      Kolkata 700 106, India.
      }\\
$^b$ {\it  Department of Physics, Visva-Bharati, Santiniketan 731235, India}
}\\
\end{center}
\vspace{.2cm}

\begin{abstract}
{\noindent \normalsize

The search for the top squark ($\lstop$) within the kinematic reach of 
Tevatron Run II is of great contemporary interest. Such a $\lstop$
can explain the baryon asymmetry of the universe
provided $120 ~GeV/c^{2} \leq \mlstop \leq m_t$. Moreover if 
$\Delta m \equiv \mlstop$ - $\mlspone$ is small, where $\lspone$ is the LSP,
the dark matter relic density as obtained from the WMAP data may 
be explained via $\lstop$- LSP coannihilation. In this scenario
the decay $\loopdk$ is likely to occur with 100 $\%$ branching ratio
but for small $\Delta m$ the conventional di-jet + $\met$ signal
becomes unobservable. We propose a new search strategy based on the 
$2j + \met$ signature accompanied by an isolated cluster of
energy which arises from a decaying heavy particle with characteristic 
decay length. Our preliminary simulations with PYTHIA indicate that 
for $100 ~GeV/c^{2} \le \mlstop \le 130  ~GeV/c^{2}$ this signal
may be observable while somewhat lager  $\mlstop$
may still provide hints of new physics.

}

\end{abstract}

\newpage
\section{Introduction} 
\label{intro4}

~~~In the  Minimal Supersymmetric Standard Model (MSSM)
\cite{susy} there are two  scalar superpartners $\stopl$ and $\stopr$, of the 
top quark which are the weak eigenstates. The mass eigenstates the 
lighter top squark ( $\lstop$ ) and the heavier top squark
($\hstop$) are linear combinations of the weak eigenstates. 
Due to mixing effects in the top squark mass matrix in the weak
basis driven by the top quark mass ($m_t$) there may be a significant
mass difference between $\lstop$ and $\hstop$. In fact the former could 
very well be the next-to-lightest supersymmetric particle (NLSP), 
the lightest neutralino ($\lspone$) being the lightest supersymmetric 
particle (LSP) by the standard assumption in R-parity conserving MSSM. 
This happens in a wide region of the MSSM parameter space. In this scenario, 
henceforth referred to as the $\lstop$ - NLSP scenario, the $\lstop$ may 
be the only strongly interacting  superpartner within the kinematic reach 
of Tevatron Run II experiments with a relatively large production 
cross-section. 

Additional interest in the light top-squark scenario stems from the 
observation that the MSSM can explain the baryon asymmetry of the universe 
via electroweak baryogenesis (EWBG) provided $120 ~GeV/c^{2} \leq \mlstop \leq m_t$ 
\cite{baryo}. The search for $\lstop$ is, therefore, a high priority program 
for the on going experiments at the Tevatron.  

The search for $\lstop$-NLSP at Tevatron Run I and LEP and, more recently, 
at Tevatron Run II produced negative results and lower bounds on $\mlstop$. 
Most of the analyses \cite{lep,tev,tev2} are based on the assumption that 
$\lstop$ decays via the Flavor Changing Neutral Current (FCNC) induced loop 
decay, $\loopdk$ \cite{hikasa} with 100 $\%$ branching ratio (BR). We also 
employ this assumption which is by and large valid if tan$\beta\geq 7$ 
where tan $\beta$ is the ratio of the vacuum expectation values for the two 
neutral Higgs bosons present in the MSSM\cite{Djouadi,shibu}. 
For lower values of this parameter the four body decay of the $\lstop$ may 
be a competing channel \cite{Djouadi,shibu,manas}

There are decay modes of the $\lstop$ - NLSP other than the above two
channels. They are the tree-level two body decay, $\lstop \ra t \lspone$ and 
the three body decay, $\lstop \ra bW \lspone$. The last modes are
kinematically forbidden for small values of the mass difference $\Delta m = \delm$  
which is the main concern of this paper. 

The search for $\lstop$-NLSP are based on the jets plus missing $E_T$ 
channel\cite{tev,tev2}. Some of the more recent works employed c-jet tagging 
by a lifetime based heavy flavour algorithm. These jets become 
softer if $\Delta m$  is 
small. As a result the efficiency of the kinematical cuts for suppressing 
the background as well as that of c-jet tagging decreases. This weakens the 
limit on $\mlstop$ from Tevatron. At Tevatron Run I the largest $\mlstop$ 
excluded was $122 ~GeV/c^{2}$ for $\mlspone = 55  ~GeV/c^{2}$. The most 
recent analysis by the D0 collaboration at 
Run II \cite{tev2} with c-jet  tagging obtained the limit $150$ 
 $~GeV/c^{2}$ for $\mlspone = 65 ~GeV/c^{2}$for the most conservative 
cross-section after including the next to leading order (NLO) corrections 
\cite{prospino}.

On the other hand
the LEP lower-bounds on $\mlstop$ are restricted mainly due to kinematics 
and are around $100 ~GeV/c^{2}$ \cite{lep}. However, much
smaller values of $\Delta m$ can be probed  in the cleaner environment 
of an $e^+ - e^-$ collider. 

The prospect of $\lstop$ - NLSP search via this decay channel at Run II 
was investigated in \cite{demina}. It was observed that a large region 
of the $\mlstop$ - $\mlspone$ parameter space corresponding to small 
$\Delta m$ is beyond the reach of Run II. For a given $\mlstop$ there
is a minimum value of $\Delta m$ that 
can yield an observable signal.

A modified strategy for $\lstop$-NLSP searches in the limit of 
small $\Delta m$ is important in its own right. The current interest in 
this search, however, is further strengthened by one of the cornerstones of the 
interface between particle physics and cosmology. A very attractive 
feature of the R-parity conserving MSSM is that the LSP ($\lspone$), is a very 
good candidate for the dark matter (DM) in the universe required, e.g, by the  
Wilkinson Microwave Anisotropy probe (WMAP) data 
\cite{WMAPdata}. The DM relic density depends on the annihilation cross-section 
(thermally averaged) of a LSP pair. The coannihilation of the LSP 
with any other supersymmetric particle(sparticle)is another
important mechanism for relic density production. This mechanism is, however, 
efficient only when the 
two coannihilating particles have approximately the same mass. Thus in the 
small $\Delta m$ scenario $\lstop$ - LSP coannihilation may indeed be an 
important mechanism for producing appropriate relic density \cite{coannistop}.

The region of the parameter space of MSSM consistent with the DM relic 
density is severely constrained by the WMAP data. Nevertheless even in more 
restricted versions of the MSSM like the minimal supergravity model(mSUGRA) 
\cite{msugra} one finds a narrow region of the parameter space where 
$\lstop$ - LSP coannihilation is an important relic density producing 
mechanism\cite{debottam}.

The search for $\lstop$-NLSP with a small $\Delta m$ is, therefore, 
important irrespective of the question of EWBG. However, it is certainly 
worthwhile to check whether $\lstop$ with mass in the quoted range 
preferred by EWBG can also produce an acceptable DM relic density. This 
was investigated in \cite{hikasa}. It was found that in a significant 
region of the allowed parameter space $\Delta m$ is indeed small 
( see figure 7 of \cite{hikasa}). 

The results of \cite{hikasa} were illustrated by specific choices of 
other MSSM parameters. In particular EWBG in the MSSM requires certain 
CP violating (CPV) phases. In a certain phase convention the relative 
phase($\phi_{\mu}$) between the higgsino mass parameter $\mu$ and the SU(2) 
gaugino mass $M_2$ is the most important one. EWBG usually requires 0.05 
$\leq \phi_{\mu} \leq$ 1. However, various uncertainties in the 
calculation does not rule out a much smaller magnitude of this phase. Thus 
calculations by neglecting this phase seems to be a reasonable 
approximation\cite{carena}.

It should, however, be emphasized that the signal proposed by us is fairly model 
independent and does not depend on the CPV phase or many of the other MSSM 
parameters at all as long as the BR($\loopdk$ )
is close to 100\%. Under this assumption the size of the  
signal depends on $\mlstop$ through the production cross-section of the 
$\lstop$ pair via the standard QCD processes and on $\mlspone$ through the 
efficiency of the kinematical cuts.

The starting point of our work is the observation that 
when mass difference $\Delta m$ is 
small, in most of the signal events, one
of the $c$-quarks from $\lstop$ pair decay is not energetic enough to produce a
jet which may pass jet selection criteria of the experiments at Tevatron.
It may be seen as an isolated energy deposit in the calorimeter coming from the
decay of a heavy particle. We call it {\em isolated cluster} ({\em IC}).
Thus the proposed signal consists of a c-jet of modest 
$E_{T}$ accompanied by missing energy and an {\em isolated cluster}.
In order to reduce the background we require another hard jet in the signal 
which in most cases comes from QCD 
radiation. Our simulations show that a set of selection criteria based on 
the above features of the signal can isolate it from  the SM background.

In this work we do not consider the prospect of fully identifying the 
flavour of the heavy, isolated, decaying object because of the rather 
small statistics. This leads to inevitable backgrounds from, e.g., 
$\bb$ events and W/Z + jets events. However, we shall analyze 
at the generator level some important characteristics of this object which 
has the potential of reducing  the SM backgrounds to a manageable level. 
At the same time we emphasize that this work is only suggestive of a new 
approach to $\lstop$ search at the Tevatron and needs detailed detector 
simulation for a  more definitive statement and that is beyond the scope 
of this work.
\begin{figure}[tb]
\begin{center}
\includegraphics[width=\textwidth]{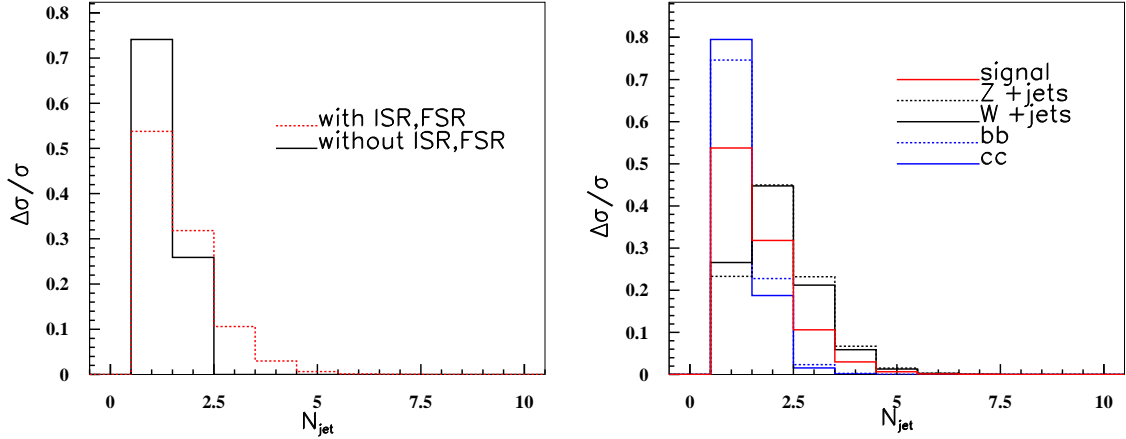}
\end{center}
\caption{The figure on left shows the distributions of jets in
         signal ($\mlstop = 120 \gev/c^{2}$, $\mlspone = 110 \gev/c^{2}$)
         with initial and final state radiation ON (dashed)
         and OFF (solid). The figure on right shows the distributions of
         jets in signal and all the major backgrounds we have analysed.} \label{fig:njet}
\end{figure}

\begin{figure}[tbp]
\begin{center}
\includegraphics[width=\textwidth]{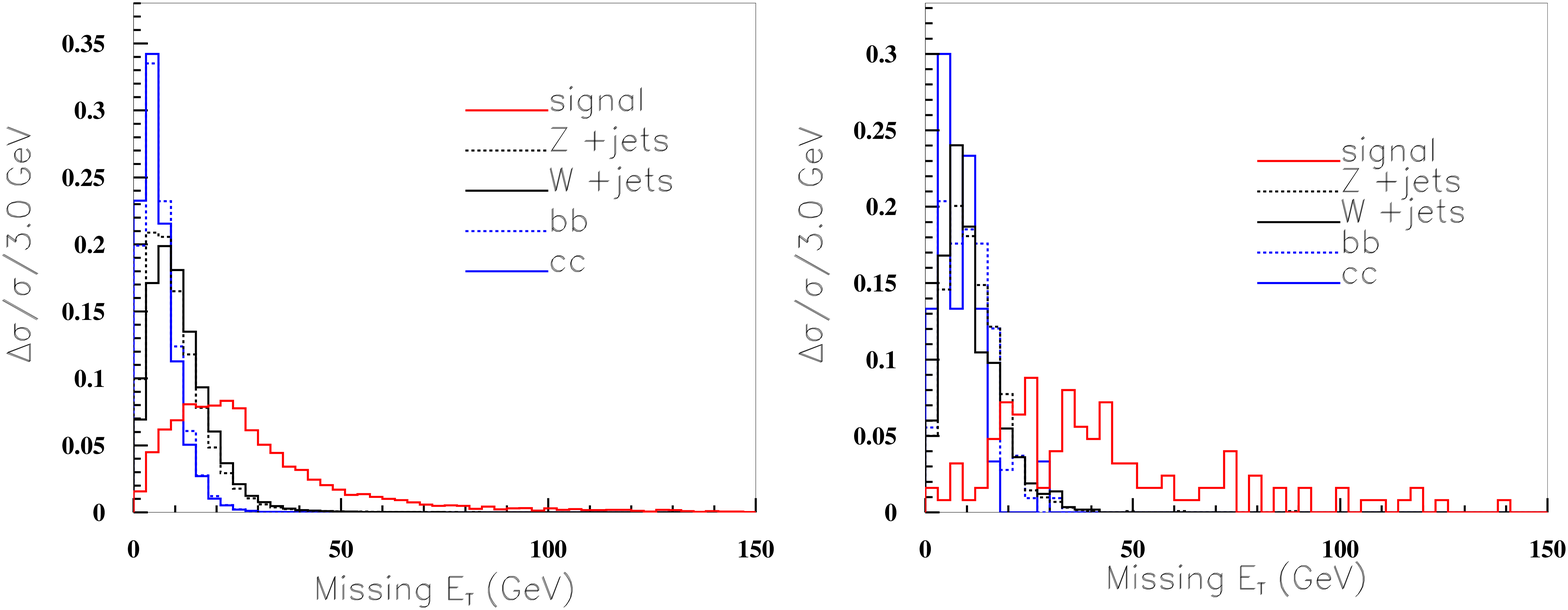}
\end{center}
\caption{The distributions of $\met $ of the signal
($\mlstop = 120 \gev/c^{2}$, $\mlspone = 110 \gev/c^{2}$) and the major
backgrounds are shown after pre-selection (left) and after all selection
cuts except the one on $\met $ (right).}
\label{fig:etmiss}
\end{figure}


We have used Pythia (v 6.206) \cite{pythia} for generation of both signal and background 
events which includes generation of the parton level events followed by 
the decay of the partons hadronization and decay of their daughters. 
Generation of both signal and background events take into account initial 
state radiation (ISR) and final state radiation (FSR). The cross-sections 
$\sigma_{\bb}$ and $\sigma_{\cc}$ are very large and most of these events 
generated with low $\surd{\hat{s}}$ are not relevant for our analysis.
To sample the $\bb$ 
and $\cc$ events better for our purpose and save computer time we have used 
a cut $\hat{p_{T}} \ge 3\gev/c$ for generation of $\bb$ and $\cc$ events
, where $\hat{p_T}$ is defined in the CM frame of the colliding partons.
We have used the toy calorimeter simulation followed by jet formation in 
Pythia (PYCELL).
\begin{enumerate}
  \item The calorimeter coverage is $\vert \eta \vert \le 3$. 
  \item A cone algorithm with 
	$\Delta R = \sqrt {\Delta\eta^2 + \Delta\phi^2}= 0.5 $ 
	has been used for jet finding with $E^{jet}_{T,min} \ge 10$ GeV 
	and $|\eta_{jet}| \le 3$ and jets have been ordered in $E_{T}$.
  \item We consider leptons $(\ell=e,\mu)$ with $E_{T}^{\ell} \ge 5 ~\gev$,  
        $\vert \eta^{\ell} \vert  \le 3$. The lepton should be isolated 
	from jets ($\Delta R (jet,\ell) \ge 0.5$).
  \item For charged particles ( $e$, $\mu$ and charged hadrons ), we have used 
	their generator level momentum as {\em track momentum} when required.
  \item For jets containing a $B$ or a $D$ hadron we have used their decay 
	length information for determining the presence of a long lived 
	particle.
\end{enumerate}

A quark or a gluon from 
(mainly) FSR is seen as a jet and in most cases this jet appears to have 
the highest $E_{T}$. This prompts us to consider a rather unusual signature 
for the signal events as mentioned below.

The backgrounds, particularly $\bb$ and $\cc$ events have very large 
cross-sections and hence we need to generate a large number of events and 
retain only a small fraction of them which pass pre-selection for detailed 
analysis. We have used the following pre-selection criteria:

\begin{enumerate}
  \item Event should have only two jets: $N_{jet} = 2$ 
        (see figure \ref{fig:njet})
  \item Events with isolated leptons are rejected.
  \item One of the jets should contain a long lived particle
        ($B$ or, $D$ hadron) and henceforth  
        called {\em matched-jet} (MJ).
  \item Event should have an {\em isolated cluster} resulting from the
        decay of a $B$ or $D$ hadron such that 
        $\Delta R (jet,{\em IC}) \ge 0.5$. The direction of the {\em isolated 
        cluster} is defined to be the direction of the decaying {\em B} or 
        {\em D} hadron. In the final selection this cluster has to be 
	identified as the signature of a long lived particle,
        the criteria for which are discussed in detail later.
  \item We assume that a {\em b-jet} with $ 30 ~GeV < E^{jet}_{T} < 50 ~GeV$ 
        is tagged with a probability $ \epsilon_{b} = 0.4$ and
        for  $E^{jet}_{T} > 50 ~GeV$,  $\epsilon_{b} = 0.5$ where 
        $\epsilon _{b}$ is the single $b$-jet tagging efficiency 
        (i.e., the ratio of the number of tagged $b$-jets and the 
        number of taggable $b$-jets).
\end{enumerate}

The pre-selection efficiency for the signal events is rather small: for 
$\mlstop = 120\gev /c^{2}, ~\mlspone = 110\gev /c^{2}$ (A) only $7.3\%$ 
events survives pre-selection; for 
$\mlstop = 120\gev /c^{2}, ~\mlspone = 105\gev /c^{2}$ (B) and 
$\mlstop = 120\gev /c^{2}, ~\mlspone = 100\gev /c^{2}$ (C) the rates are
$9.7\%$ and $9.6\%$ respectively. For $\tt$, $\bb$, $\cc$ and
{\em V + jet} events ({\em V = W, Z}) rates are $0.03\%$, $0.41\%$, $0.19\%$, 
$1\%$ and $2\%$  respectively (see table \ref{tab:single} for details). 
The leading order cross-section at $Q = \mstop$, where $Q$ is the QCD
scale for $\lstop \lstop^*$ production is from \cite{Beenakker}.
All background cross-sections have been computed by CalcHEP (version 2.3.7)
\cite{calchep} at $Q = \sqrt{\hat s}$.
The largest background comes from {\em Z+jets} production.
The variation of the cross-section of this process with the
QCD scale is not very severe.

\begin{table}[tbp]
\begin{center}
\begin{tabular}{cc|ccccccc|c}
\hline\hline
Process  & $\sigma$ (pb)  & \multicolumn{7}{c|}{Efficiencies for selection cuts}  & $\mathcal{L} . \sigma . \epsilon$ \\ \cline{3-9}
         &        & Presel  & Cut 1   & Cut 2   & Cut 3   & Cut 4  & Cut 5  & Final($\epsilon$) &\\
\hline
         &        &         &         &         &         &        &        &         &        \\
signal A & 4.2    & 0.0734   & 0.0236   & 0.0155   & 0.0154   & 0.0225  & 0.0489  & 0.00056 & 18.8   \\
signal B & 4.2    & 0.0966   & 0.0325   & 0.0209   & 0.0234   & 0.0338  & 0.0527  & 0.00059 & 19.8   \\
signal C & 4.2    & 0.0959   & 0.0362   & 0.0239   & 0.0276   & 0.0346  & 0.0442  & 0.00057 & 19.2   \\
$ZZ$     & 1.006  & 0.0095   & 0.0049   & 0.0029   & 0.0037   & 0.0030  & 0.0032  & 6.4E-05 &  0.52  \\
$WZ$     & 2.39   & 0.0035   & 0.0016   & 0.0009   & 0.0013   & 0.0010  & 0.0012  & 1.2E-05 &  0.23  \\
$WW$     & 8.76   & 0.0020   & 0.0010   & 6.2E-05  & 0.0006   & 0.0003  & 0.0008  &  0      &  0     \\
$\tt$    & 3.82   & 0.0003   & 0.0003   & 0.0002   & 0.0002   & 8.7E-05 & 5.5E-05 & 2.0E-06 &  0.06  \\
$\cc$    & 7.8 E07& 0.0019   & 8.4E-05  & 2.0E-07  & 0.0004   & 0.0002  & 0.0015  &  0      &  0     \\
$\bb$    & 1.6 E07& 0.0054   & 0.0003   & 2.8E-07  & 0.0017   & 0.0015  & 0.0007  &  0      &  0     \\

{\em W+jets} & 8.7 E03& 0.0091  & 0.0041  & 5.0E-05 & 0.0027  & 0.0011 & 0.0033  &  0      &  0     \\
{\em Z+jets} & 3.0 E03& 0.0189   & 0.0043  & 0.0001  & 0.0048  & 0.0043 & 0.0070 & 4.0E-07 &  9.6 \\
         &        &         &         &         &         &        &       &          &        \\
\hline\hline
\end{tabular}
\end{center}
\caption{Efficiency table for the signal and SM backgrounds to pass 
	 various selection criteria. Column 3 shows the efficiency for 
	 pre-selection; columns 4-8 show the efficiencies for each cut 
	 combined with the effect of pre-selection ; column 9 shows the
         efficiency for final selection. (see text for details)
	 The last column shows the expected number of events to pass all 
	 selection(rejection) criteria for ${\mathcal L} = 8 ~fb^{-1}$.} 
\label{tab:single}
\end{table}
\begin{table}[tb]
\begin{center}
\begin{tabular}{p{25mm}p{10mm}p{10mm}p{10mm}p{10mm}p{10mm}p{10mm}p{10mm}p{10mm}p{10mm}p{10mm}}
\hline\hline
 $\mlstop$        &  \multicolumn{2}{c}{100}  &\multicolumn{2}{c}{110}  & \multicolumn{2}{c}{120} & \multicolumn{2}{c}{130} & \multicolumn{2}{c}{140} \\
 $\sigma$ (pb)    &  \multicolumn{2}{c}{11.2} &\multicolumn{2}{c}{7.5}  & \multicolumn{2}{c}{4.2} & \multicolumn{2}{c}{3.3} & \multicolumn{2}{c}{2.25} \\
\hline
                  &  $N_{sig}$ & $S$ & $N_{sig}$ & $S$ & $N_{sig}$ & $S$ & $N_{sig}$ & $S$ & $N_{sig}$ & $S$\\
\hline
 $\Delta m = 10 $ &  64.5  & 20.0 & 32.4 & 10.0 & 18.8 & 5.8  & 17.8 & 5.5 & 11.9 & 3.7  \\
 $\Delta m = 15 $ &  68.1  & 21.1 & 45.0 & 13.9 & 19.8 & 6.1  & 17.5 & 5.4 & 13.1 & 4.1  \\
 $\Delta m = 20 $ &  61.8  & 19.1 & 35.4 & 11.0 & 19.2 & 5.9  & 17.2 & 5.3 & 13.9 & 4.3   \\
\hline\hline
\end{tabular}
\end{center}
\caption{Production cross-sections for different $\mlstop$ are given. Signal
         events surviving ($N_{sig} = \mathcal{L} . \sigma . \epsilon$) for
         ${\mathcal L} = 8 ~fb^{-1}$ for different values of $\Delta m$
         are shown alongwith the corresponding significance
         $S$. The mass parameters are in units
         of $\gev/c^{2}$.}
\label{tab:signi}
\end{table}

\begin{figure}[!htb]
\begin{center}
\includegraphics[width=0.7\textwidth]{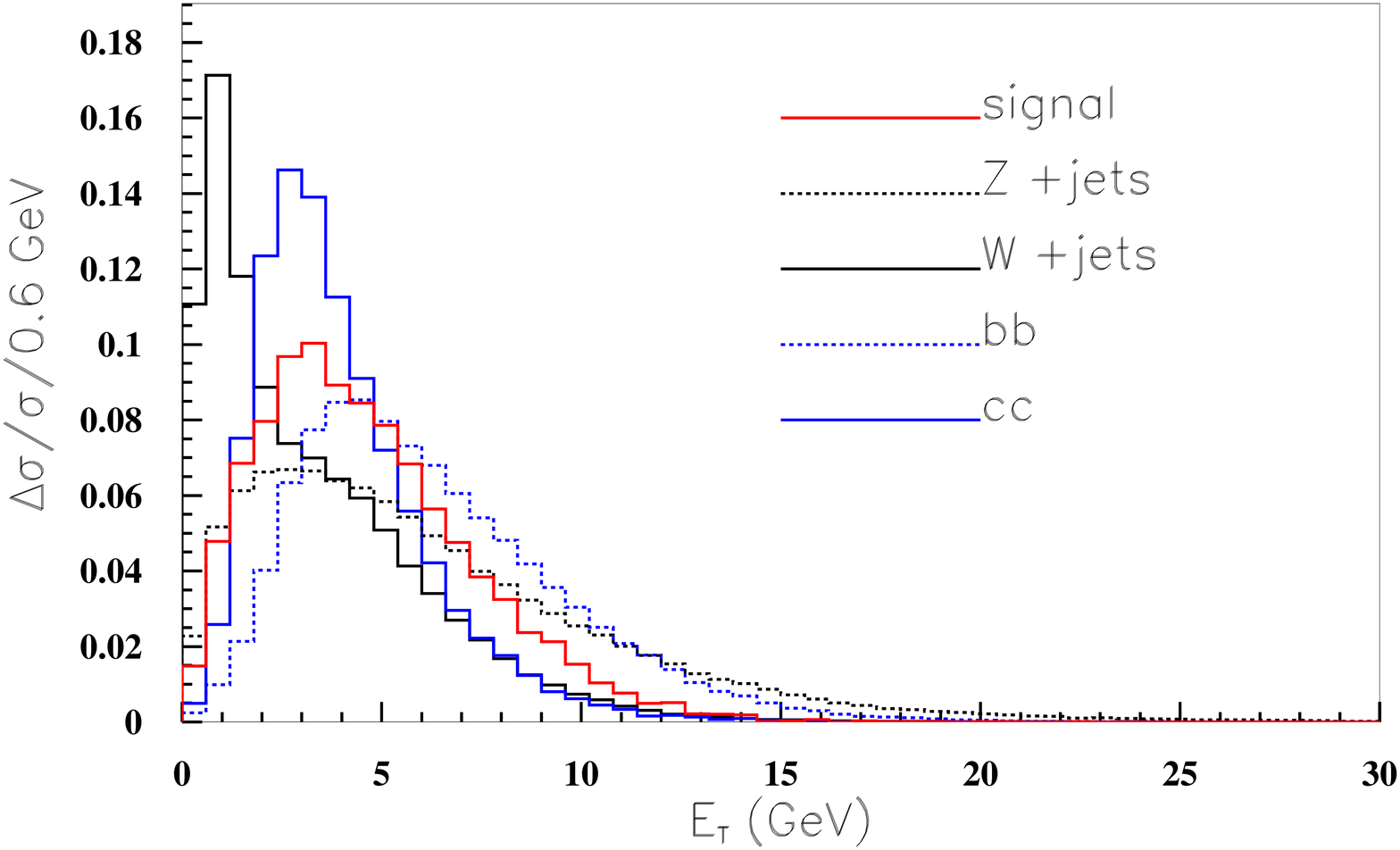}
\end{center}
\caption{The distributions of $E_{T}$ of the {\em isolated cluster} are
         shown for the signal
         ($\mlstop = 120 \gev/c^{2}$, $\mlspone = 110 \gev/c^{2}$)
         and the major backgrounds.}
\label{fig:iso_pt}
\end{figure}
\begin{figure}[tb]
\begin{center}
\includegraphics[width=0.7\textwidth]{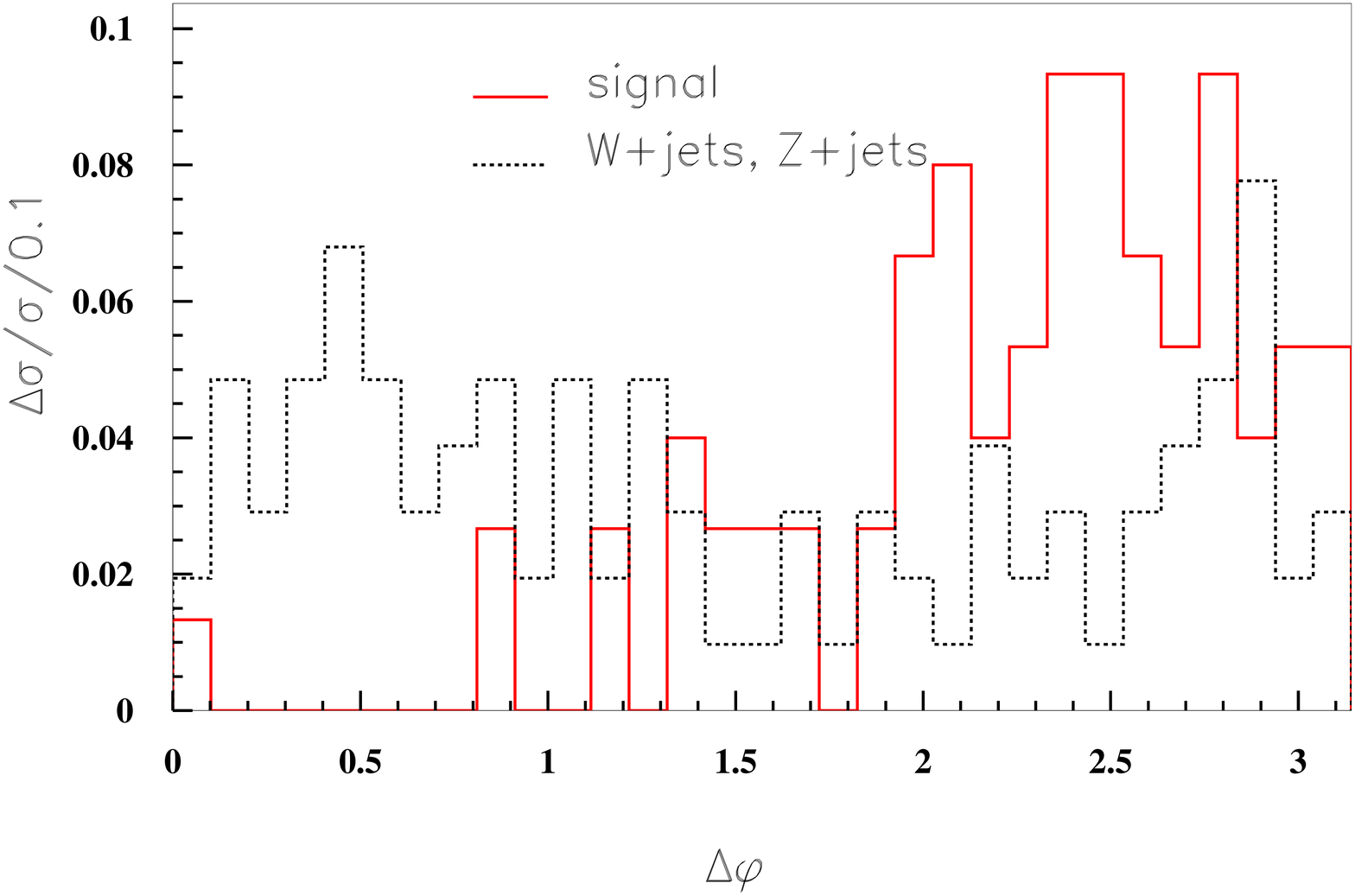}
\end{center}
\caption{ The distributions of $ \Del \phi (IC, MJ)$ (in radians) are shown
          for the signal ($\mlstop = 120 \gev/c^{2}$,
          $\mlspone = 110 \gev/c^{2}$) and the {\em W+jets} and {\em Z+jets}
          backgrounds after all cuts except Cut 5 (see text).}
\label{fig:phi_diff}
\end{figure}
\begin{figure}[tb]
\begin{center}
\includegraphics[width=0.7\textwidth]{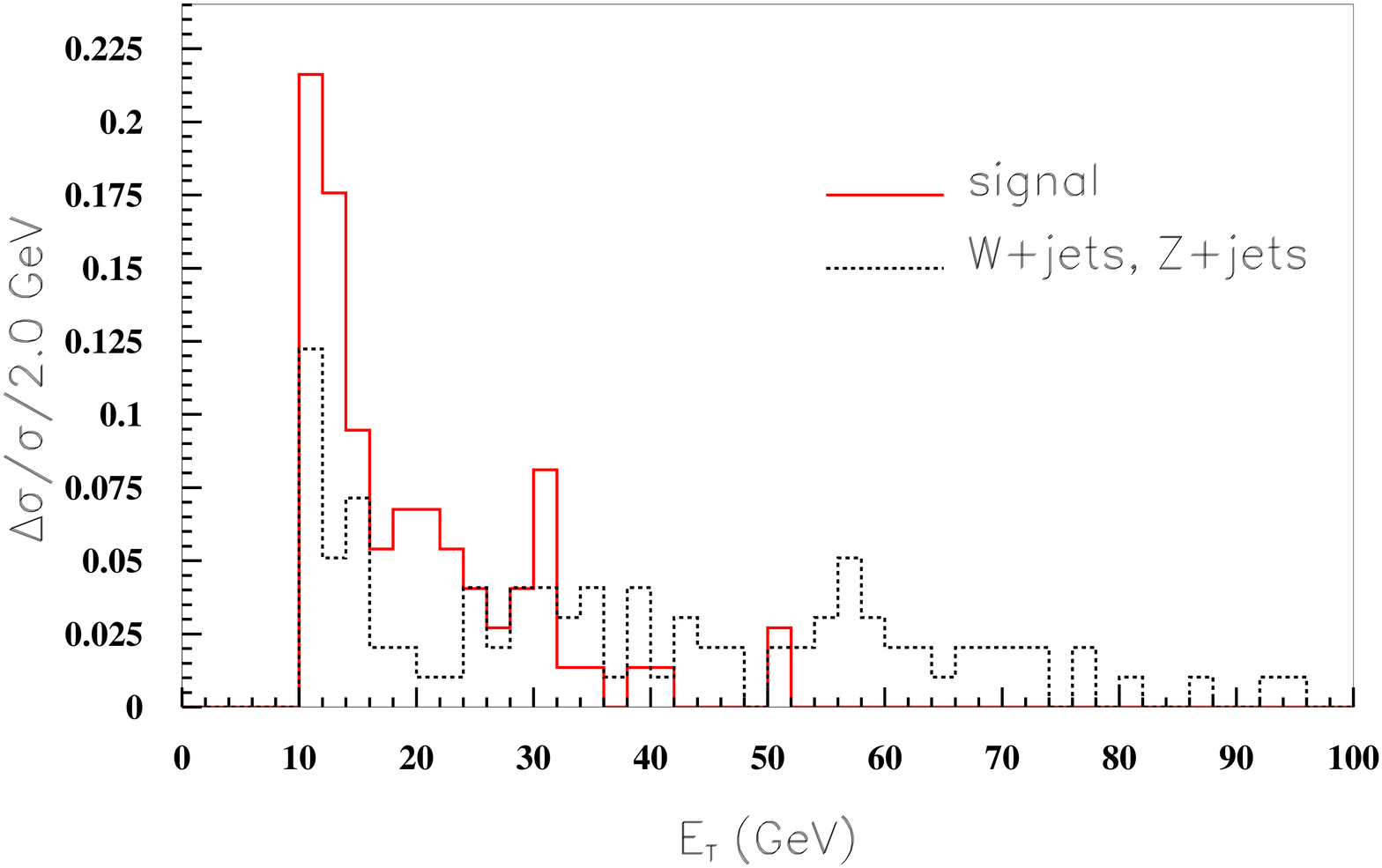}
\end{center}
\caption{The distributions of $E_{T}^{MJ}$ of the signal
         ($\mlstop = 120 \gev/c^{2}$, $\mlspone = 110 \gev/c^{2}$)
         and the {\em W+jets} and {\em Z+jets} backgrounds after
         all cuts except Cut 5 (see text).}
\label{fig:pt_matched}
\end{figure}

\begin{figure}[tb]
\begin{center}
\includegraphics[width=\textwidth]{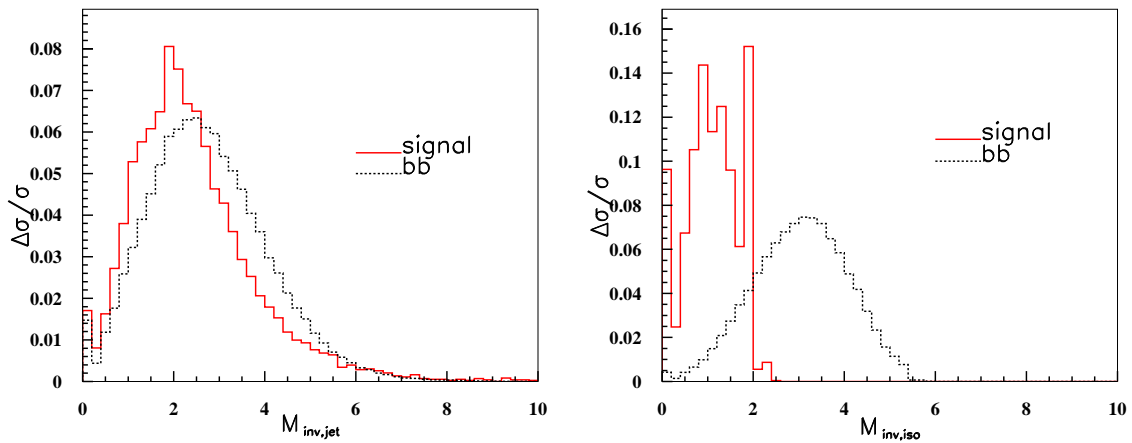}
\end{center}
\caption{The figure shows the distributions of $M_{inv}^{MJ}$ (left) 
         and $M_{inv}^{IC}$ (right) reconstructed using charged
         tracks associated with the {\em matched-jet} and 
         the {\em isolated  cluster} in signal
         ($\mlstop = 120 \gev/c^{2}$, $\mlspone = 110 \gev/c^{2}$)
          and $\bb$ events after pre-selection (see text) 
	 .}
\label{fig:mass}
\end{figure}

For final selection (rejection) of signal (background) events we demand 
the following:

\begin{enumerate}
  \item For the first jet we require $E^{jet1}_{T} \ge 25 ~GeV$ and 
	$\vert\eta_{jet1}\vert \le 1.5$ and for the second jet 
        $\vert\eta_{jet2}\vert \le 1.5$ (Cut 1).
  \item Events should have $\met > 40 \gev$ (Cut 2) (see figure \ref{fig:etmiss})
  \item The long lived particle in the {\em matched-jet} should have decay 
        length $\ge 1.5$ mm. (Cut 3)
  \item	The {\em isolated cluster} should be central and have a 
	minimum $E_{T}$: $\vert\eta^{IC}\vert \le 1.5$ and 
	$E_{T}^{IC} \ge 5 ~GeV$. It should have a decay length 
	$\ge 0.1 ~mm$. (Cut 4) (see figure \ref{fig:iso_pt})  
  \item In the signal we expect the {\em isolated cluster} and 
        the {\em matched-jet} to be approximately back-to-back in the 
	transverse plane. So, the cut $\Del\phi({\em IC},{\em MJ}) > 85^{\circ}$ 
	(see figure \ref{fig:phi_diff} ) rejects background, 
	particularly {\em W+jets}  and {\em Z+jets} events. The 
        {\em matched-jet} is most likely to be the leading jet in {\em W+jets} 
        and {\em Z+jets} events whereas it is the 2nd jet in the signal events. 
	We therefore select events whose  $E^{\em MJ}_{T} < 40 ~GeV$ 
	(see figure \ref{fig:pt_matched}).

	We also partially reconstruct invariant mass of the {\em matched-jet} 
        ( $M_{inv}^{MJ}$ ) using the charged tracks associated with it.
        Similarly $M_{inv}^{IC}$ is reconstructed for the 
	{\em isolated cluster}. The cuts $M_{inv}^{MJ} \le 4.5 \gev/c^{2}$ and
        $M_{inv}^{IC} \le 2 \gev/c^{2}$, reject the $\bb$ events
        and also reduce {\em W+jets} and {\em Z+jets} backgrounds 
	(see figure \ref{fig:mass}). The combined effect of these cuts (Cut 5) 
	for the signal and the dominant backgrounds is shown in column 8 of 
	table \ref{tab:single}.
\end{enumerate}

In table \ref{tab:single} the selection efficiencies for each cut (1 - 5) 
includes the effects of pre-selection. The eighth column shows the 
efficiency for the final selection.

Although a very high rejection factor of ($10^{-7}$) is achieved for the 
$\bb$ events and no event in the simulated sample survives, this may still be
dangerous as $\sigma_{\bb}$ is very large. Since the signal events do not 
have spectacular signatures, it is not possible to apply more stringent 
criterion on any of the features and still retain a good signal. 
It may be required to exploit the subtle features of the {\em matched-jet} and the 
{\em isolated cluster} to get rid of the $\bb$ events. There are a few 
features which may be exploited to this end although it may be experimentally 
challenging:
\begin{itemize}
  \item Number of charged tracks associated with the {\em isolated cluster}.
  \item Upper cut on the lifetime related observables for the {\em matched-jet}, 
	expected to be a {\em c-jet} and the {\em isolated cluster}, expected 
        to come from the decay of a $D$ hadron in the signal.
  \item presence of a $K^{\pm}$ in the {\em isolated cluster} which carries a 
	significant fraction of its $p_{T}$. 
  \item More reliable reconstruction of the invariant masses using full
        detector information.
\end{itemize}

Since these observables are rather {\em delicate}, they may be simulated only 
using detailed detector modelling and estimation of the effects of criteria 
based on these observables is beyond the scope of this analysis. However it 
may be said with some degree of confidence that $\bb$ and other background 
events may be  further suppressed by using judicious choice of such criteria.

Our final results are presented in table \ref{tab:signi}. In view of the LEP 
limits \cite{lep} we have looked into signals with 
$\mlstop \ge 100 ~GeV/c^{2}$ and we have chosen 
$\delm$ (in $GeV/c^{2}$) = 10, 15 and 20. The significance is defined as 
$S = N_{sig}/\sqrt{N_{bkg}}$ where $N_{sig}$ ( $N_{bkg}$ ) is the number of 
signal ( background ) events passing the selection criteria for an integrated 
luminosity of ${\mathcal L} = 8 ~fb^{-1}$. For $\mlstop \le 130 ~GeV/c^{2}$
we may hope for a {\em discovery} while beyond that it may be restricted to
the level of a {\em hint}. The parameter space probed by us contains a part of 
the region interesting in the context of EWBG. 

Again, it should be emphasized that this very preliminary analysis is designed 
to provoke the experimentalists to scan the region interesting from the point 
of view of dark matter relic density and EWBG. 

\newpage
{\bf Acknowledgment}:
AD, MM and NB acknowledge financial support from  Department of Science and Technology,
Government of India under the project  No (SR/S2/HEP-18/2003).
MM also acknowledges support from  Department of Science and Technology,
Government of India under the project  No (SP/S2/K-25/96-VI).
A part of this work was done when AD and NB were in the
Department of Physics, Jadavpur Universiry, Kolkata 700 032, India.





\begin{thebibliography}{99}
\markright{Bibliography}

\bibitem{susy}
For reviews on Supersymmetry, see, {\it e.g.},
H. P. Nilles, \PREP (110,1984,1);
H. E. Haber and G. Kane, \PREP (117,1985,75) ;
J. Wess and J. Bagger, {\it Supersymmetry and Supergravity}, 2nd ed.,
(Princeton, 1991); M. Drees, P. Roy and R. M. Godbole, {\it Theory and
Phenomenology of Sparticles},
(World Scientific, Singapore, 2005).

\bibitem{baryo}
M. Carena, M. Quiros and C.E. Wagner, \PLB(380,1996,81),
hep-ph/9603420; \NPB (503,1997,387), hep-ph/9702409;
\NPB (524,1998,3),hep-ph/9710401; D.Delepine, J.M. Gerard,
R.Gonzalez Felipe and J.Weyers, \PLB(386,1996,183),hep-ph/960440;
J. McDonald, \PLB(413,1997,30),hep-ph/9707290; J.M. Cline and G.D. Moore,
\PRL(81,1998,3315),hep-ph/9806354.
\bibitem{lep} LEPSUSYWG, ALEPH, DELPHI, L3 and OPAL collaborations,
note LEPSUSYWG/04-02.1 (http:\// lepsusy.web.cern.ch/lepsusy/Welcome.html)


\bibitem{tev} CDF Collaboration: T. Affolder {\em et~al.},
\PRL(88,2002,041801); T. Altonen {\em et~al.},\PRD(76,2007,072010)
(arXiv 0707.2567).




\bibitem{tev2}D0 collaboration: V. M. Abazov et al,\PRL(93,2004,011801);
V. M. Abazov et al \PLB(645,2007,119); V. M. Abazov
 arXiv:0803.2263.


\bibitem{hikasa} K. Hikasa and M. Kobayashi, \PRD(36,1987,724).


\bibitem{Djouadi} C. Boehm, A. Djouadi and Y. Mambrini, \PRD(61,2000,095006).


\bibitem{shibu} S. P. Das, A. Datta and M. Guchait, \PRD(65,2002,095006).


\bibitem{manas} S.P.Das, Amitava Datta and Manas Maity, \PLB(596,2004,293);
S.P. Das, \PRD (73,2006,115004).


\bibitem{prospino} W. Beenakker et al,\NPB (515,1998,3).


\bibitem{demina} R. Demina et al \PRD (62,2000,035011)

\bibitem{WMAPdata}
  D.~N.~Spergel {\it et al.},
  arXiv:astro-ph/0603449.

\bibitem{coannistop}
C. Boehm, A. Djouadi and
Manuel Drees,
\PRD(62,2000,035012);
J.~R.~Ellis, K.~A.~Olive and Y.~Santoso,
\ASTR(18,2003,395)
                                                                                
[arXiv:hep-ph/0208178].

                                                                                                                             
\bibitem{msugra}
A.~H.~Chamseddine, R.~Arnowitt and P.~Nath,
\PRL(49,1982,970);
R.~Barbieri, S.~Ferrara and C.~A.~Savoy,
\PRL(119,1982,343);
L.~J.~Hall, J.~Lykken and S.~Weinberg,
\PRD(27,1983,2359);
P.~Nath, R.~Arnowitt and A.~H.~Chamseddine,
\NPB(227,1983,121);
N. Ohta, \PTP( 70,1983,542).
                                                                                                                             
                                                                                                                             
\bibitem{debottam}
Utpal Chattapadhyay, Debottam Das, Amitava Datta and Sujoy Poddar,
\PRD(76,2007,055008).
                                                                                



\bibitem{carena} C.Balazas, M. Carena and C.E.M. Wagner, \PRD (70,2004,015007)
V. Cirigliano, S. Profamo, M.J Ramsey Musolf
\JHEP(067,2006,002)

\bibitem{pythia} T. Sjostrand, P. Eden, C. Friberg, L. Lonnblad, G. Miu,
S. Mrenna and  E. Norrbin,\CPC (135,2001,238);
For a more recent version see,
\JHEP (0605,2006, 026)

\bibitem{Beenakker}W. Beenakker, M. Kramer, Tilman Plehn and Michael Spira
 arXiv : hep-ph/9810290
                                                                                                                             


\bibitem{calchep} See,e.g., A.Pukhov, CalcHEP$-$a package for
evaluation of Feynman diagrams and integration over multi-particle phase space
(hep-ph/9908288).
For the more recent versions see: http://www.ifh.de/pukhov/calchep.html.

\end{thebibliography}
\end{document}